\documentclass[sigconf,review=false]{acmart}

\setcopyright{none}
\renewcommand\footnotetextcopyrightpermission[1]{}
\usepackage{booktabs}
\usepackage{graphicx}
\usepackage{balance}
\usepackage{stfloats}

\begin{document}
\title{MDwAIstScheduler: Bringing On-Device Voice Documentation into Clinical Practice}

\author{Diego Mardian}
\affiliation{%
  \institution{Arizona State University}
  \city{Tempe}
  \state{Arizona}
  \country{USA}}
\email{dmardia2@asu.edu}

\author{Frank Liu}
\affiliation{%
  \institution{Arizona State University}
  \city{Tempe}
  \state{Arizona}
  \country{USA}}
\email{fwliu1@asu.edu}

\begin{abstract}
Clinical documentation forces physicians to split attention between the patient and their keyboard, and much of it spills into uncompensated after-hours work. We present MDwAIstScheduler, a low-cost, belt-worn pipeline that lets a physician speak naturally during the encounter and have the resulting medications, allergies, labs/orders/referrals, follow-up scheduling, vitals, and problems land in the EHR as review-ready drafts. Building on our earlier prototype, which relied on cloud speech recognition and a cloud language model, the current pipeline runs both transcription and intent extraction \emph{entirely on-device}. Using a medical-domain automatic speech recognition (ASR) model and a 1.7B-parameter language model we fine-tuned for clinical action extraction, no patient audio or text leaves the device, and the structured drafts are written directly into the Elation EHR for the physician to confirm. The result is a documentation tool that removes keyboard work from the visit without removing the clinician from the record, allowing them to focus on what matters most, patient care, while reducing burden at the same time.
\end{abstract}

\maketitle

\begin{figure*}[b]
  \centering
  \vspace{15pt}
  \includegraphics[width=\linewidth]{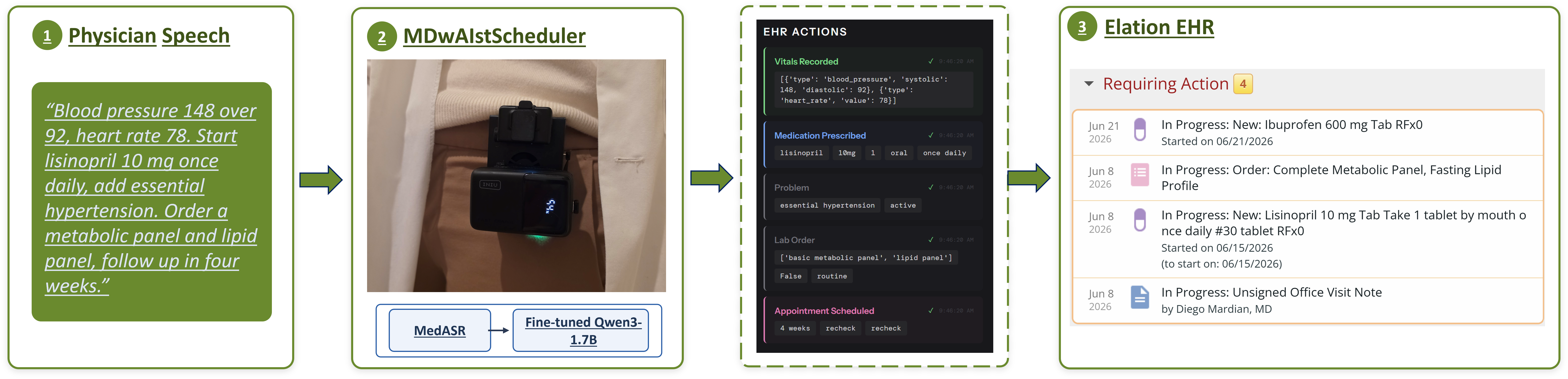}
  \caption{End-to-end MDwAIstScheduler workflow: (1) the physician speaks naturally; (2) the belt-worn device transcribes and extracts structured actions entirely on-device; (3) only structured entries are written to Elation as drafts for physician sign-off.}
  \label{fig:system}
\end{figure*}
\section{Introduction}
For a practicing physician, documentation is one of the most expensive parts of the day. In ambulatory practice, clinicians spend only about 27\% of their time in direct contact with patients and 49.2\% on the EHR and desk work, with additional documentation pushed into personal time after clinic~\cite{sinsky2016}; this load is a major contributor to burnout, which affects over 43\% of U.S. physicians~\cite{olson2025}. The ways clinicians cope today each carry a cost. Typing or clicking into the EHR during the visit pulls gaze and attention away from the patient and measurably degrades communication and engagement~\cite{alkureishi2016, asan2014}. Deferring documentation to the end of the day preserves the encounter but shifts the work into after-hours documentation; recent national EHR-log studies find physicians spend well over an hour per day working in the EHR outside scheduled patient time~\cite{holmgren2024}. Commercial ambient AI scribes reduce typing by listening to the whole visit, but they generate narrative notes rather than discrete chart actions, and they stream patient audio to the cloud, which raises privacy and cost barriers for smaller practices~\cite{tierney2024, ghatnekar2025}.

MDwAIstScheduler targets this problem as an end-to-end pipeline rather than a single tool: the physician speaks naturally during the encounter, and the system delivers structured, review-ready entries into the chart without any keyboard interaction and without the patient's words leaving the room. The value to the provider is concrete. Eye contact and conversation are preserved because there is no screen or wrist device to look at; the discrete actions a visit actually produces (a prescription, a lab order, a problem, a follow-up) are captured as they are spoken instead of being reconstructed later; and because everything runs locally, the privacy and infrastructure barriers that keep cloud scribes out of small and rural clinics largely disappear.

We made these gains by further developing our earlier prototype, which demonstrated hands-free \emph{scheduling} using cloud speech recognition and a cloud language model~\cite{mdwaist_v1}. The present pipeline improves on it in two ways that matter for real deployment. First, \textbf{both} the speech recognizer and the language model now run \textbf{on the device itself}, so raw audio and transcripts never go over a network, directly addressing the privacy concern that blocks most ambient tools from many clinical settings. Second, the pipeline no longer just handles scheduling: it documents six categories of clinical action (medications, allergies, labs/orders/referrals, follow-up scheduling, vitals, and problems) and writes them into a production EHR (Elation) as drafts. This abstract focuses on the pipeline as a whole and how it behaves in practice (Figure~\ref{fig:system}).

\section{The System in Practice}
The system follows a single, simple path from speech to chart, shown end-to-end in Figure~\ref{fig:system}.

\paragraph{(1) Physician speech.} Consider a routine hypertension follow-up. The physician, mid-conversation and without touching a keyboard, says: \emph{``Alright, blood pressure today is 148 over 92, heart rate 78. Let's start lisinopril 10 milligrams once daily for the hypertension and add a new diagnosis of essential hypertension. Also order a basic metabolic panel and a lipid panel, and let's get her back in four weeks to recheck.''}

\paragraph{(2) On-device processing.} The belt-worn device captures the utterance and processes it entirely locally: MedASR transcribes the audio, and the fine-tuned Qwen3-1.7B model converts the transcript into structured clinical actions. Importantly, no audio or text ever leaves the device, only the resulting structured fields do. In Figure~\ref{fig:system}, it shows this exact encounter, in which a single spoken sentence is decomposed into its separate documentation items, correctly extracted from the natural speech.

\paragraph{(3) Drafts in the EHR.} The structured actions are written to the patient's Elation chart as \emph{drafts}, where they appear in the ``Requiring Action'' queue (Figure~\ref{fig:system}, right). This draft-first design keeps a clinician in the loop for every entry: the device proposes, the physician disposes, and nothing reaches the record without human sign-off. Together with on-device inference, this is what turns the ambient AI from a compliance vulnerability into something a small practice can actually adopt. After the visit, the physician opens the chart, sees the drafts pre-populated, and signs or edits them, turning the full document for that visit into seconds of review.

\subsection{Accuracy}
A model small enough to run on the device must still extract clinical actions accurately. On a held-out set of 98 clinician-style utterances spanning all action categories, our fine-tuned 1.7B model produced valid JSON 100\% of the time and passed a clinical-adequacy judge---whether a downstream reader would extract the correct action---on 98.0\% of cases, exceeding an un-tuned model nearly five times its size (8B, 85.7\%) while running about $4.7\times$ faster. Fine-tuning is what makes this possible: judge-pass on the same 1.7B backbone jumps from 48.0\% before tuning to 98.0\% after. On the speech side, our medical-domain recognizer (MedASR) gets the highest clinical-fidelity score of the models we tested while running at roughly $40\times$ lower latency than the general-purpose models it beats, because its errors avoid the drug names and clinical terms that matter most. In short, the on-device stack is accurate enough for draft-then-confirm use, while still running in near real-time.

\section{Future Work}
Our most important next step is a \textbf{live field study with practicing physicians} to measure the system's actual impact: time saved per encounter, change in documentation completed during versus after visits, perceived administrative burden, and effect on patient-perceived attentiveness. Benchmarks establish that the model is accurate, however only a deployment can establish that the workflow is actually faster and less burdensome in the messy reality of a clinic.

The same pattern of unobtrusive capture, on-device processing, structured draft, and human confirmation also generalizes beyond its clinical use-case. Any hands-busy professional who must narrate structured data into a system of record fits the design, such as other clinical roles (nursing, EMS, dentistry, veterinary), non-clinical fields (field service, lab work, aviation checklists), and other systems of record (different EHRs, legal or insurance documentation). Because inference is fully local, the system also suits rural and low-resource settings with no reliable connectivity or per-query cloud cost. Beyond the field study, we plan to add multi-turn dialogue for disambiguating ambiguous commands, broaden the action vocabulary, and extend EHR write-back toward physician-confirmed orders and assistive billing-code suggestion, always preserving human sign-off.

\section{Conclusion}
MDwAIstScheduler shows that a fully on-device, belt-worn voice assistant can turn natural clinical speech into structured, review-ready EHR documentation without sending patient data to the cloud. A compact fine-tuned model runs locally but extracts multiple structured actions from a single utterance, and a draft-first design keeps a clinician in control of every entry.

\balance

\end{document}